\documentclass[aps,prl,twocolumn,superscriptaddress]{revtex4-2}

\usepackage{graphicx}
\usepackage{verbatim}
\usepackage{relsize}
\usepackage{mathrsfs}
\usepackage{color}
\usepackage{colortbl}
\usepackage{ulem}
\usepackage{blindtext}
\usepackage{lipsum}
\usepackage{hyperref} 
\usepackage{multirow}
\usepackage{newtxtext}                                        
\usepackage{booktabs} 
\usepackage[dvipsnames,table]{xcolor}
\usepackage{hyperref}
\usepackage[textwidth=1.5cm,textsize=footnotesize]{todonotes}
\usepackage{amsmath,amsfonts,amssymb}
\usepackage{float}
\usepackage{wrapfig}
\usepackage{array}

\newcommand{\tabitem}{~~\llap{\textbullet}~~}

\begin{document}

\title{Topology and geometry under the nonlinear electromagnetic spotlight}
\author{Qiong Ma}\affiliation{Department of Physics, Massachusetts Institute of Technology}\affiliation{Department of Physics, Boston College}  
\author{Adolfo G. Grushin}
\affiliation{Institut Ne\'el, CNRS and Universit\'e Grenoble Alpes, Grenoble, France}
\author{Kenneth S. Burch}\affiliation{Department of Physics, Boston College} 
\date{\today}

\maketitle

\section{abstract}
\textbf{For many materials, a precise knowledge of their dispersion spectra is insufficient to predict their ordered phases and physical responses. Instead, these materials are classified by the geometrical and topological properties of their wavefunctions. A key challenge is to identify and implement experiments that probe or control these quantum properties. In this review, we describe recent progress in this direction, focusing on nonlinear electromagnetic responses that arise directly from quantum geometry and topology. We give an overview of the field by discussing new theoretical ideas, groundbreaking experiments, and the novel materials that drive them. We conclude by discussing how these techniques can be combined with new device architectures to uncover, probe, and ultimately control novel quantum phases with emergent topological and correlated properties.
}

\section{I. Introduction}
Nonlinear electromagnetics in quantum materials provides the means to uncover, probe, and ultimately control new phases of matter (Fig.~\ref{fig:overview}). In general the field has been propelled by new materials, theoretical developments and novel experimental techniques. For example, the first prediction of a nonlinear electromagnetic response by M. Goeppert-Mayer in 1929 relied on the simultaneous absorption of two photons whose energies were half those of the electronic transition involved~\cite{meyer31}. The confirmation of two-photon absorption would take thirty years, relying on the invention of the laser with its high power, coherence, and narrow wavelength. 

Widely tunable lasers and new detectors resulted from optimizing nonlinear responses in materials, such as second-, third- and high-harmonic, difference-frequency and nonlinear photocurrent generation~\cite{LiuNonlinDetect,boyd2020nonlinear,SturmanBook}. These nonlinear electromagnetic phenomena now serve to detect broken symmetries, new phases of matter, and the examination of quantum geometry and topology of electron wavefunctions~\cite{Harter2017SHG,Xu2020TiSe2,Sun2019SHG,mciver2020light}. Nearly simultaneously, numerous theoretical and experimental efforts have focused on using nonlinear approaches to strengthen Cooper pairing (via control of phonons and magnons), reshape the band structure (Floquet engineering), and control fundamental excitations~\cite{TOka2019FloqRev,Sie2019WTe2,Moessner:2017ur,basov2017towards}. 

The desire to reveal, understand and tune the wavefunction geometry and topology is driven by the recent discovery and engineering of quantum materials \cite{basov2017towards, wen2019choreographed, armitage2018weyl,BurchNatMag18}. Quantum spin liquids, topological semimetals, and superconductors are examples of new electronic and magnetic phases that have shifted the focus from symmetry breaking to quantum geometry. In many of these phases nonlocality, superposition and entanglement are fundamental to their properties, and are not revealed by classical probes. Uncovering and investigating the underlying physics of these emergent phases requires new experimental means and theoretical ideas to explore the quantum geometry and topology of their wavefunctions. The complexity of real materials poses additional challenges when the desired phases do not emerge or the responses include trivial contributions. To overcome material limitations, efforts have shifted to manipulating samples with engineered heterostructures and via nonlinear techniques. The latter also has the advantage of potentially controlling the emergent quasiparticles for quantum computation.

In this review we will outline the current theoretical understanding of nonlinear responses, and their connection to quantum geometry and topology. A summary of possible effects and the ranges at which they emerge can be found in Fig.~\ref{fig:Possible}. We also discuss how the realization of these ideas required emerging materials and fabrication techniques that produced an array of new topological phases. After discussing these, we give an overview of recent experimental progress in probing the quantum geometry and topology of these materials, ending with a discussion of results in correlated materials and future directions. Though not discussed in detail, we note caution should be taken in the interpretation and implementation of nonlinear responses, as factors other than symmetry and quantum geometry can contribute. The theoretical aspects of these effects are discussed in Box 1, while the experimental complications are summarized in Box 2. In addition, for those new to quantum materials we have provided a summary of the relevant bulk materials in Box 3, and 2D crystals in Box 4.

\section{II. Overview of non-linear responses \label{sec:overview}}
\par Nonlinear electrodynamics refers to material response functions that depend on second- or higher-order powers of the electromagnetic field. As these are typically smaller than linear responses, they require higher intensities. Additionally, as they involve multiple fields, nonlinear response functions are more constrained by crystal symmetries than linear responses. These constraints are used to expose symmetry breaking and geometric properties of the wavefunction that can uncover, probe and control new phases of matter. 

To describe nonlinear responses we consider an incident light beam, and assume it acts continuously in time (a continuous wave) with a wavelength $\lambda$ that we consider to be much larger than the sample size ($2\pi/\lambda \equiv |\mathbf{q}| \to 0$). With these assumptions we exclude complications due to short pulses or spatially inhomogeneous beams, discussed in Box 1. Furthermore, we leave for Box 1 extrinsic effects due to electronic scattering.  With these simplifications, the response of a medium that generates a time-dependent photocurrent $j_i(t)$ can be expressed in powers of an incident electric field $E_i(t)$ as~\cite{Sipe00}
%%%
\begin{widetext}
\begin{equation}
    \label{eq:Js}
    j_i(t) = \sum_{j}\sigma^{(1)}_{ij}E_j(t) + \sum_{jk}\sigma^{(2)}_{ijk} E_{j}E_{k}(t)+ \sum_{jkl}\sigma^{(3)}_{ijkl} E_{j}E_{k}E_{l}(t)+\cdots,
\end{equation}
\end{widetext}
%%%
The responses proportional to even powers of the electric field ($E^{2n}$, $n=1,2,...$) are only allowed when the lattice structure breaks inversion symmetry because $E^{2n}$ is even under inversion, while $j_i$ is odd. In contrast, terms proportional to odd powers of electric fields are always symmetry-allowed. In addition, lattice symmetries constrain which components of $\sigma^{(i)}$ are non-zero, set by the space-group of the material ~\cite{boyd2020nonlinear, SturmanBook}. 

The goal of this section is to describe to what extent non-linear responses $\sigma^{(i>1)}$ are determined by local and global properties of the wavefunction. A local property is defined by how the wavefunction $\left| \psi^{n}_\mathbf{k} \right\rangle$ for a given electronic band $n$, varies locally as we change its momentum $\mathbf{k}$ in the Brillouin zone. In contrast, global properties are defined by comparing the wavefunction before and after it is transported around a closed path, or closed surface, in $\mathbf{k}$ space. Global properties do not depend on the path taken, and thus do not depend on local details or energetics of the Hamiltonian. This ultimately  quantizes responses to an integer in units of fundamental constants, which is topological because it can only be changed by a global, rather than local, perturbation of the Hamiltonian. 

We begin by connecting local properties of the wavefunction to the expectation value of electrical currents. Specifically we consider the basic quantum mechanical principle by which the position is mapped to derivatives in momentum space. With this principle, $\mathrm{A}_{nm}(\mathbf{k}) = \left\langle\psi^{n}_\mathbf{k}\right| i\boldsymbol{\nabla}_\mathbf{k} \left| \psi^{m}_\mathbf{k} \right\rangle$, known as the Berry connection, is related to the expectation value of the position operator written in momentum space~\cite{Sipe00}. The expectation value of the position operator determines the charge polarization, and therefore its time derivatives determine the expectation values of currents. A response that depends on $\mathrm{A}_{nm}$ is geometrical since gradients of the wavefunction offer a sense of how the wavefunction varies locally in $\mathbf{k}$ space. This quantity and its derivatives determine nonlinear responses~\cite{Morimoto:2016iu,Sipe00}.

Responses that probe global properties of the wavefunction are far more elusive than those that probe local properties. Global responses are special because they are defined by momentum integrals over a closed surface in momentum space, e.g. the Brillouin zone, of gauge-invariant geometrical properties of the electronic wavefunction. An important  gauge-invariant geometrical property of a band is $\boldsymbol{\Omega}^{n}_\mathbf{k}=\boldsymbol{\nabla}_\mathbf{k} \times \mathrm{A}_{nn}(\mathbf{k})$, known as the Berry curvature. It is invariant under changes in the phase of the wavefunction, much like a magnetic field is invariant under the choice of different electromagnetic gauges. Its flux over a closed momentum surface is a global robust property, a topological invariant called the Chern number $C$. Observables that depend on $C$ are topological. For example, in two-dimensional insulators, the Hall conductivity $\sigma_H$ is a linear response determined by the integral of $\boldsymbol{\Omega}^{n}_\mathbf{k}$ over the Brillouin zone, and it is quantized to $\sigma_H = Ce^2/h$. Below we discuss how, under certain approximations, there are also quantized topological nonlinear responses (in units of $e^3/h^2$)~\cite{de2017quantized}, and the challenges encountered in measuring them~\cite{de2017quantized,chang2017unconventional,rees2019quantized,Ni2020b}.\\

\textbf{Second-order responses: photogalvanic effects and second-harmonic generation}

The second-order term in Eq.~\eqref{eq:Js}, determined by $\sigma^{(2)}$, is especially significant since it governs responses that are known to be connected to different geometrical properties of the wavefunction. For periodically oscillating fields with frequency $\omega$ the two Fourier components $\pm\omega$ can interfere constructively, generating a response oscillating at $2\omega$ known as second-harmonic generation (SHG). If they interfere destructively the result is a non-oscillating DC response known as the intrinsic photogalvanic effect or the bulk photovoltaic effect (BPVE) (Fig.~\ref{fig:Possible}\textbf{a}). The attributes \textit{intrinsic} and \textit{bulk} remind us that the direction of the photocurrent is set by crystal symmetry and the bandstructure, rather than by extrinsic factors, such as a preferential direction at the interface in engineered heterostructures~\cite{boyd2020nonlinear, SturmanBook}, or scattering effects discussed in Box 1.

The intrinsic photogalvanic effect is explicitly determined by geometrical properties of the wavefunction~\cite{Morimoto:2016iu,tokura2018nonreciprocal,ahn2020low}. It can be split into two contributions: a time-independent photocurrent, referred to as a shift current, and the injection current, a non-oscillating current which grows linearly in time~\cite{boyd2020nonlinear,Sipe00}. It is also practical to separate photocurrent responses due to linear and circular polarization, termed the linear and circular photogalvanic effects, or LPGE and CPGE, respectively. The presence of time-reversal symmetry determines which polarization generates shift and injection currents. For non-magnetic materials, in which time-reversal symmetry is present, the LPGE and the CPGE are determined by the shift and injection currents, respectively. More generally, when time-reversal symmetry is broken, such as in magnetic materials, both types of photocurrents can occur with any polarization~\cite{Holder19,deJuan:2020jm}. 

The shift current can be linked to the polarization difference between the excited electron and hole. This response can be expressed as the difference between the expectation value of Berry connections (the average real space position of each state) of the photoexcited pair~\cite{Fregoso:2017ex}. The name shift current originates from the possibility to write the DC photocurrent response to linear polarization as a vector that represents the real-space shift of the photoexcited electron and hole~\cite{Sipe00}. Device dimension, wavefunction geometry and band structure can all limit the shift current output~\cite{Tan:2019da,patankar2018resonance,osterhoudt2019colossal}.
% Scattering is also crucial in preventing the electron-hole from recombining and producing an equal and opposite shift current, resulting in near zero response~\cite{Sotome2019Shift}. 
Finding the conditions for its enhancement to aid material searches is a challenging question in general, but it is possible with simple models. For example, in two-band models, the skewness of the polarization distribution (determined by the third moment of the polarization) constrains the magnitude of the shift current through a sum rule~\cite{patankar2018resonance}. 

Unlike the shift current, the injection current can directly probe the band-resolved Berry curvature. The circular photocurrent generated by the Dirac surface states of three-dimensional time-reversal invariant topological insulators is an important example.  Under circular polarization the optical selection rules vary around the surface in momentum space defined by optically allowed transitions ($S_\mathbf{k}$). In the two-band limit, relevant to the surface bands, the transition probability is the Berry curvature~\cite{Hosur:2011gfa,Kim:2017dh}. In three-dimensions the trace of the injection current tensor can be topologically quantized because it is determined by the Berry curvature flux over $S_\mathbf{k}$~\cite{de2017quantized}. For quantization to occur $S_\mathbf{k}$ must be closed, and enclose a single topologically protected two-band crossing, known as a Weyl node. In real materials however, Weyl nodes come in pairs, acting as a Berry curvature monopole-antimonopole pair. If the nodes are separated in energy, $S_\mathbf{k}$ can enclose one node only, quantizing the injection current to an integer equal to the topological charge of the monopole $C$, times $\pi e^3/h^2$~\cite{de2017quantized}. In Weyl semimetals with mirror symmetries, like TaAs, nodes of opposite charge occur at the same energy, leading to a zero injection current trace.
In chiral Weyl semimetals, those that lack all mirror symmetries, the nodes are separated in energy, opening a window of frequencies (the size of which depends the material) for which the injection current trace is quantized. This prediction was then extended to topological nodes involving more than two bands~\cite{chang2017unconventional,deJuan:2020jm}, in a class of chiral metals known as multifold semimetals~\cite{Bradlyn2016,chang2017unconventional,tang2017multiple,chang2018topological}.

% , those that lack all mirror symmetries, this implies a quantized CPGE~\cite{de2017quantized}, because their topologically protected two-band crossings, known as Weyl nodes, are not degenerate in energy (unlike in mirror symmetric Weyl semimetals like TaAs). Each node act as Berry curvature monopoles, and if they are separted in energy the surface of optically allowed transitions encloses one node. The CPGE in this case is exactly quantized to an integer equal to the topological charge of the monopole, times $\pi e^3/h^2$~\cite{de2017quantized}. This prediction was then extended to topological degeneracies involving more than two bands~\cite{chang2017unconventional,Flicker2018Chiral,deJuan:2020jm}, in materials known as multifold semimetals~\cite{Bradlyn2016,chang2017unconventional,tang2017multiple,chang2018topological}. However, the additional bands that reduces the energy window where quantization is observable, compared to Weyl fermions~\cite{Flicker2018Chiral}. 

Second-harmonic generation is also determined by geometric properties of the wavefunction~\cite{Tan:2019da,patankar2018resonance}. For two-band models, the shift current and second-harmonic generation are related~\cite{patankar2018resonance}, implying that the skewness of the polarization distribution also restricts the SHG magnitude in these models. Finding general principles, valid for any number of bands, remains an open question.

Second-harmonic generation and the photogalvanic effect are the monochromatic limit of more general phenomena. When a material is subject to two oscillating electric fields with frequencies $\omega_1$ and $\omega_2$, their interference causes responses that oscillate at a frequency $\omega_{-}=\omega_1-\omega_2$, known as difference-frequency generation (or optical parametric amplification), $2\omega_1$ and $2\omega_2$, their respective second harmonics, and at $\omega_{+}=\omega_1+\omega_2$, known as sum-frequency generation (see Fig.~\ref{fig:Possible}\textbf{b}). Difference-frequency generation near the monochromatic limit, when $\omega_{+}\gg \omega_{-}\gg \tau^{-1}$, with $\tau$ a typical scattering time, reduces to two response tensors that coincide with the injection and shift current tensors. These oscillate out of phase and can be separated~\cite{deJuan:2020jm}. The geometrical content of sum-frequency generation remain to be explored in depth, although general expressions exist~\cite{Sipe00}.

Second-order nonlinear responses have different phenomenology for metals and insulators~\cite{Genkin68}. Both allow for interband effects, where an electron-hole pair is excited with a frequency related to the incoming probe, but a finite Fermi surface allows metals to also display intraband effects. Most of them can be accounted for semiclassically (although exceptions exist~\cite{deJuan:2020jm}), by considering the motion of a wave packet around the Fermi surface subject to the external electric fields~\cite{sodemann2015quantum,Morimoto2016,Golub2020,Glazov:2014cp}. An important example is the nonlinear Hall effect proportional to the Berry curvature dipole $\partial_{k_{a}}\Omega_b$ (Fig.~\ref{fig:Berry2D}\textbf{a})~\cite{Genkin68,sodemann2015quantum}. It can be traced back to the anomalous velocity proportional to the Berry curvature, and it appears in systems with a dipolar distribution of Berry curvature in momentum space. Even when the total net Berry curvature is zero, the Berry curvature dipole leads to non-zero net Berry curvature in the current-carrying state, and therefore generates a Hall voltage as a second-order response~\cite{moore2010confinement,sodemann2015quantum}. This current is the $\omega\tau\to0$ limit of the intraband LPGE, when we pick the particular component where current is transverse to the electric field. A short overview of extrinsic effects is included in Box 1.\\

\textbf{Higher-order responses: third- and higher-harmonic generation} 

Third-order terms in Eq.~\eqref{eq:Js} have been also related to geometrical properties in the context of topological semimetals and two-dimensional (2D) materials~\cite{Parker2019diag,Cheng19b}. Third-harmonic generation, by which three electric fields oscillating at frequency $\omega$ interfere constructively to generate a pulse at $3\omega$, has been related to the Berry curvature and its quadrupole moment in topological semimetals~\cite{Parker2019diag}. It is unknown if the third-order effect known as the Jerk current~\cite{Fregoso:2018kq}, a current increasing quadratically in time, is tied to topological responses. 

Lastly, high-harmonic generation, studied for decades in atomic and molecular gases, is gaining focus as a property of quantum materials. High-harmonic generation can be thought as the result of an intense electric field enabling a charge to coherently oscillate many times to produce higher frequency light. The geometrical properties~\cite{Morimoto:2016iu,Lee19,Yoshikawa2017HHG,Luu2018HHG}, as well as emergent quasiparticles (e.g., spinons and doublons)~\cite{Murakami2018} are predicted to enter higher-harmonic generation.  In nodal metals high-harmonic generation may be enhanced by topological properties~\cite{Lee19}, and end states of one-dimensional insulators can enhance the high-harmonic yield dramatically~\cite{Bauer18}. High-harmonic generation has also been proposed as a tool to observe two-dimensional topological phase transitions~\cite{Chacon18}, and to characterize surfaces of three-dimensional topological insulators~\cite{Jia19}.

\section{III. Experimental toolbox for probing nonlinear responses}
The optical SHG, which measures 2$\omega$ response due to incident probe at $\omega$, is the most common technique to detect second-order responses. It was developed to study the symmetry breaking properties at surfaces or interfaces \cite{shen1989surface} and the photon energies typically dealt with are in the near infrared and optical range. It has also been recently applied to study various 2D materials to probe their crystallographic symmetry \cite{li2013probing, Sun2019SHG}. Particularly, the SHG shows a resonant effect and optical selection rules in WSe$_2$ when 2$\omega$ matches with its exciton emission energy \cite{seyler2015electrical}. Such resonance-type experiments can reveal microscopic details beyond symmetry properties. They are feasible because the incident photon energies lie close to the band gap in these materials either at $\omega$ or $2\omega$. 

Topological materials usually feature narrow band gaps (< 100 meV, such as in topological insulators \cite{hasan2010colloquium}) or gapless electronic structures (such as Dirac or Weyl semimetals \cite{armitage2018weyl}), with topological and Berry curvature properties strongly confined to the band edges, possibly coexisting with topologically trivial bands close by in energy. Therefore, in order to sensitively and exclusively probe quantum geometry, one needs to perform SHG measurements in the infrared and THz regimes, which is technically challenging due to the lack of intense light sources, sensitive detectors, and frequency-selective components. To this end, Patankar et al.~\cite {patankar2018resonance} performed SHG measurements in TaAs with the incident photon energy as low as 0.5 eV (see Fig.~\ref{fig:3Dmat}\textbf{d}). 
%while Ni et al.\cite{Ni2020b} measured CPGE down to 0.2 eV.

An alternative approach to move down to low energies is nonlinear photocurrent generation. Experimentally, such measurements do not need highly sensitive detectors and optical filters since the probed signal is a DC current. Such currents are routinely measured with great precision using electrical contacts or through free-space THz emission if excited with a pulsed laser. To this end, infrared and THz excitations have been successfully applied to probe a range of topological semimetals~\cite{ma2017direct, osterhoudt2019colossal, xu2018electrically, ma2019nonlinear, ji2020photocurrent,Ni2020b}. The generation of intrinsic DC current upon light excitation is essentially a second-order process (the monochromatic limit of difference frequency generation $\omega - \omega$), as explained in Section II. Depending on its microscopic mechanism and light polarization, the photocurrent can directly probe momentum space integrals over the Berry curvature or Berry connection and carry geometric information about the electronic wavefunction~\cite{ji2020photocurrent,osterhoudt2019colossal,ma2017direct}.  

Recent theoretical advances identified a means to move to even lower energies, closer to the Fermi surface. Specifically, second-order responses from the quantum geometry of the wavefunction can manifest as an intraband SHG response in metallic systems, which can be measured electrically~\cite{sodemann2015quantum,isobe2020high} (see Fig.~\ref{fig:Possible}\textbf{a}). In electrical SHG measurements, one injects electrical current at frequency $\omega$ (from the DC limit to microwaves) and selectively measures second-order (or higher-order) electrical responses. Such low-frequency electrical excitations usually induce intraband transitions and are therefore sensitive to electronic states near the Fermi surface. By employing electrical gates, the Fermi level can be further tuned, enabling energy sensitivity in this all electrical version of SHG. Indeed this was recently demonstrated via a nonlinear anomalous Hall effect from the  Berry dipole~\cite{ma2019observation, kang2019nonlinear}.  \\

\section{IV. Recent Experimental Advances}
\noindent \textbf{Nonlinear responses in topological insulators}

Topological insulators have macroscopic properties that are rooted in the topological nature of the wavefunction~\cite{hasan2010colloquium}, most notably the emergence of protected edge or surface states. Since most topological insulators are inversion symmetric (centrosymmetric), their inversion breaking surface or edge states can dominate nonlinear responses. They are easily grown as thin films by molecular-beam epitaxy (MBE), or mechanically exfoliated into thin flakes from bulk crystals, and subsequently fabricated into various device geometries with gate tunability.

Topological insulators provided an important test case for nonlinear probes, demonstrating their ability to reveal nontrivial Berry curvature and the complexities involved in doing so. Hosur predicted that a topological insulator surface state can generate a CPGE and a photon-drag effect related to its nontrivial Berry curvature~\cite{hosur2011circular}. However, the generation of photocurrent requires a careful consideration of the spin texture. Namely, since the spin points in the plane of a typical topological insulator surface state, one primarily excites one side of the cone if the circularly polarized light has an angular momentum component in-plane. For normal incident light this will not be the case and thus one should not observe a CPGE. It was noted that the application of strain or in-plane magnetic field could break the symmetry and enable a CPGE \cite{hosur2011circular}. 

An alternative approach was taken by several groups, namely exciting with light at a finite angle of incidence, such that the angular momentum of the photon contained an in-plane component. Specifically, the CPGE in Bi$_{2}$Se$_{3}$ measured along the \textit{y} direction, only occurred when the beam was at a finite angle of incidence in the \textit{xz}- plane, consistent with the spins in the surface states being locked perpendicular to the electron momentum \cite{GedikBSPHotoconductivity, 2014NatSR4E4889D, Pan2017CPGE}. In these initial measurements the role of the topological surface states was not entirely clear as the photon energies ($1.17 \rightarrow 1.55$ eV) were far beyond the band gaps of the materials ($\approx 0.3$ eV). 

Several works aimed to enhance the surface contribution by tuning the Fermi level via Sb doping \cite{2016PhRvB93h1403O}, altering the surface electric field with an ionic liquid \cite{2014NatSR4E4889D}, and studying the combined effect of gating (tuning Fermi level) and photon energy (changing the optical transitions involved) \cite{Pan2017CPGE}. In general, it was found that the topological surfaces are crucial, though one must consider the role of the bulk bands as the excitation energies were typically above the band gap. Specifically, since the bulk states are also spin-split they can contribute to the relative absorption of right versus left circularly polarized light and thus the CPGE. Nonetheless, the response was generally maximized when the Fermi level lies at the surface Dirac point where the mobility is also largest, which is consistent with CPGE being limited by scattering (Fig.~\ref{fig:3Dmat}\textbf{a}). However, the precise location in gate voltage of the peak and sign of the CPGE were found to depend sensitively on photon energy. Such dependence results from tuning from transitions primarily between the valence band and the surface states to transitions between the surface states and the conduction band. Interestingly, the sign of the nonlinear response can also serve as probe of phase transitions from normal to topological insulators \cite{2016PhRvL.116w7402T}.

To entirely focus on the surface response, it is natural to move to lower photon energies. One group employed ferromagnetic Cr$_{0.3}$(Bi$_{0.22}$Sb$_{0.78}$)$_{1.7}$Te$_{3}$ films, with the Fermi level tuned in the Zeeman gap of the surface states \cite{Ogawa2016}. Photocurrents in the mid-IR (below the band gap) were only observed in the ferromagnetic state. Consistent with theoretical predictions~\cite{Hosur:2011gfa}, the CPGE required applying an in-plane magnetic field. Crucial to this response is the spin-momentum locking that controls the $k$-space location of absorption of a given helicity of light. In addition, this locking leads to the Dirac cone moving with applied field. When combined, these two effects meant the photocurrent followed the magnetization in magnitude and direction. In contrast, applying an out of plane magnetic field opened a Zeeman gap, resulting in nearly zero photocurrent.
 
\vspace{2mm}

\noindent \textbf{Nonlinear responses in Weyl semimetals}

Inversion-breaking Weyl semimetals allow for the generation of second-order responses which in turn provided insights into their novel quantum geometry (Box 3). CPGE has been so far observed as an interband process in Weyl semimetals, of which the understanding begins with optical selection rules with circularly polarized light. Due to angular momentum conservation, circularly polarized light leads to asymmetric optical transition probabilities within one Weyl cone \cite{chan2017photocurrents}. Such asymmetry gives rise to non-zero current. However, the Weyl cones always appear in pairs with opposite chirality separated in $k$-space, and in materials with mirror symmetry they occur at the same energy. Under the excitation of light of the same handedness, the two cones, if they are upright (no tilting), generate currents of the same magnitude but in opposite directions, resulting in a zero net photocurrent. Such a cancellation is avoided if the Weyl cones are tilted, which naturally occurs when the Weyl nodes are not at high-symmetry points (such as TRIM) \cite{chan2017photocurrents,Konig17}. When the Weyl cone is titled, the optical excitation connects electronic states at different energies on different sides of the cone. With the chemical potential away from the Weyl point, Pauli blocking serves as an additional selection rule, preventing excitation from one side of the Weyl cone. The combination of optical selection rules and Pauli-blocking allowed for non-zero photocurrent generation \cite{chan2017photocurrents,zhang2018photogalvanic}. 

To exclusively excite within Weyl cones and avoid transitions to higher energy bands,  infrared continuous-wave light from a CO$_2$ laser source (photon energy $\sim$ 120 meV) was employed and a large CPGE was observed that is finite only in the crystal symmetry-allowed configuration, demonstrating its intrinsic nature \cite{ma2017direct}. The direction of the current was used to infer the chirality configuration of Weyl nodes in momentum space. In addition to CPGE, another group observed significant shift current response in TaAs also with a CO$_2$ laser,  arising from large change in charge polarization when electrons are excited from below to above the Weyl nodes (Fig.~\ref{fig:3Dmat}\textbf{b})~\cite{osterhoudt2019colossal}. This was enabled by fabricating relatively small devices and using symmetry to separate thermal from nonlinear responses. The shift current in TaAs was found to be an order of magnitude larger than ever previously observed. This was in good agreement with theoretical calculations, that confirmed its Weyl node origin. 

A series of optical SHG measurements were also performed in TaAs with excitation energy from 0.5 eV to 1.5 eV \cite{Wu2017giant,patankar2018resonance}.  TaAs revealed an extremely large second-harmonic response at room temperature (orders of magnitude larger than GaAs) with a resonance at excitation of 0.7 eV~\cite{patankar2018resonance} (Fig.~\ref{fig:3Dmat}\textbf{d}). The giant response is closely related to the polar nature of the crystal and theoretical advances formulated it through  the ``skewness" (third cumulant) of the polarization distribution function. Due to the relatively high photon energy, non-Weyl bands are primarily excited in the measurements and contribute to the response.  

Turning to the time domain, photocurrent and SHG were studied in TaAs using ultrafast techniques \cite{sirica2019tracking,gao2020chiral}. If one uses an ultrafast laser pulse to drive the photocurrent, it decays on a sub-picosecond time scale and radiates THz into free space that can be directly measured without electrodes (Fig.~\ref{fig:3Dmat}\textbf{c}). In TaAs, ultrafast photocurrent was generated by a 1.55 eV excitation. They revealed the coexistence of CPGE flowing in the \textit{ab}-plane and shift current flowing along the polar \textit{c}-axis \cite{sirica2019tracking}, consistent with what is expected by symmetry and had previously observed by steady-state photocurrent measurements~\cite{ma2017direct, osterhoudt2019colossal}. The same group later showed that the ultrafast photocurrent could break the crystal symmetry, resulting in additional SHG response originally hidden at equilibrium \cite{sirica2020photocurrent}. Importantly, it was found that such transient symmetry breaking can only be observed when the SHG probe energy (fundamental frequency) is in resonance with the pump photon energy, suggesting its electronic origin.  The observation was then attributed to changes in the spatial distribution of electron wavefunctions due to the generation of shift current. This experiment suggests that photocurrent generation can potentially break the crystal symmetry and drive electronic transitions on ultrafast timescales.

Nonlinear photocurrents have also been observed in Type II Weyl semimetals, including $T_\mathrm{d}$- Mo(W)Te$_2$ \cite{ji2019spatially} and TaIrTe$_4$ \cite{ma2019nonlinear}. These materials are all layered crystals and can, therefore, be easily fabricated into thin flake devices with different electrode configurations.  Among those studies, the spatially-varying CPGE in MoTe$_2$ is particularly interesting \cite{ji2019spatially}. MoTe$_2$ undergoes a phase transition from $T^\prime$ (centrosymmetric) to $T_\mathrm{d}$ (non-centrosymmetric Weyl phase) at around 250 K \cite{armitage2018weyl}. In the $T_\mathrm{d}$ phase, an out-of-plane screw rotational symmetry (C$_{2\nu}$) remains and forbids the generation of in-plane CPGE with normal incident light. Against such symmetry analysis, CPGE at normal incidence was measured below 250 K. However, using electrodes in a circular pattern, it was found that the CPGE is rotating in space around the center of the laser beam \cite{ji2019spatially}. Such rotation was understood from the real space field gradient due to the Gaussian shape of the beam, which leads to asymmetric carrier excitation and relaxation dynamics.  Wavelength-dependent measurements from visible to mid-infrared showed a significantly enhanced  response at low photon energies,  suggesting its Berry curvature origin due to the presence of Weyl nodes. This same approach was later utilized to detect the orbital momentum of light \cite{ji2020photocurrent}.

\vspace{2mm}
\noindent \textbf{Nonlinear responses in atomically thin materials}

One may wonder whether the approaches taken to study surfaces of topological insulators are also applicable to monolayer graphene, given the similarity in their dispersion (the existence of a Dirac cone). However due to time reversal and inversion symmetry, the Berry curvature is zero everywhere except at the node. Nonetheless, as described in Box 1, the linear dispersion alone naturally leads to strong odd-order nonlinearities.\cite{Glazov:2014cp} Indeed, there are numerous demonstrations of strong nonlinear responses resulting from semiclassical dynamics in the THz frequency range, including third-harmonic generation~\cite{Soavi2018THG,Hendry2010THG} and the photon-drag effect~\cite{Glazov:2014cp}. At higher photon energies, the interband responses require a full quantum mechanical approach, revealing resonances when $n\hbar\omega=2E_\mathrm{F}$ ($n$ is an integer number)~\cite{Mikhailov2014THG}. Indeed, experiments show extremely strong nonlinear responses that can be tuned by photon energy and gating, including third-harmonic responses~\cite{Soavi2018THG, Hendry2010THG} and high-harmonic responses~\cite{Yoshikawa2017HHG}. 

The effect of symmetry breaking and Berry curvature on nonlinear responses can be directly seen in a comparison between monolayer graphene and monolayer MoS$_{2}$~\cite{Yoshikawa2017HHG, Liu2017HHG} (Box 4). In the latter case, the broken inversion allows for even-harmonic generation, which in MoS$_{2}$ was primarily polarized perpendicular to the incident laser. This was a consequence of the large Berry curvature, producing an anomolous velocity and thus perpendicular polarization to the high-harmonic generation~\cite{Liu2017HHG}. A similar effect was observed by comparing amorphous SiO$_{2}$ with inversion symmetry to crystalline, inversion broken quartz. For the SiO$_{2}$ experiments the use of high energy photons and careful alignment of the polarization to crystal axes allowed for the reconstruction of the Berry curvature throughout the Brillouin zone~\cite{Luu2018HHG}. Beyond the Berry curvature, the high-harmonic generation in MoS$_{2}$ revealed the role of correlations, via enhancement of the odd-harmonic responses with reduced thickness, due to increased Coulomb interactions~\cite{Liu2017HHG}. Furthermore, the spectrum of high-harmonic generation could also reveal fractionalization, allowing the measurement of the dynamics of doublons and holons in Mott insulators~\cite{Murakami2018}.

The broken inversion in MoS$_{2}$, as with hBN, in fact generates a valley-contrasting Berry curvature (Fig.~\ref{fig:Berry2D}\textbf{b}, also see Section II). Such unique Berry curvature distribution has already given rise to many interesting linear responses, such as circularly polarized photoluminescence and the valley Hall effect~\cite{xu2014spin}. Moreover, due to its non-centrosymmetric nature, the valley-contrasting Berry curvature may be important to understand and formulate strong even-order nonlinear responses, including optical second-harmonic responses as already observed in 2$H$-TMDC~\cite{li2013probing, seyler2015electrical} and hBN~\cite{li2013probing}, and electrical second-order responses as proposed by Isobe et al.~\cite{isobe2020high} (Fig.~\ref{fig:Berry2D}\textbf{e-f}). Interestingly, breaking the inversion symmetry in graphene (through coupling with substrate or applying perpendicular electric field) can also produce a valley-contrasting Berry curvature~\cite{Glazov:2014cp}. 

Lastly, a Berry dipole can exist in a wide range of non-centrosymmetric 3D lattices, including Weyl semimetals~\cite{moore2010confinement, sodemann2015quantum} (Box 3). In real experiments, second-order electrical responses could have other origins, such as Joule heating and accidental symmetry breaking due to the extrinsic effects. To demonstrate its Berry curvature origin, the device engineering, proper material choice and electrical tuning of 2D materials were crucial. Symmetry considerations limit the number of 2D materials with an in-plane polar direction (Box 4). With these restrictions in mind, thin layer $T_\mathrm{d}$ W(Mo)Te$_2$ and strained 2$H$-TMDC demonstrated the nonlinear Hall effect~\cite{zhou2020highly,ma2019observation,kang2019nonlinear} (Fig.~\ref{fig:Berry2D}\textbf{c}). In addition to the intraband process, the Berry curvature dipole can also give rise to in-plane CPGE with normal incidence through an interband process of optical excitation, as demonstrated in monolayer WTe$_2$~\cite{xu2018electrically} (Fig.~\ref{fig:Berry2D}\textbf{d}).  

\vspace{2mm}
\noindent \textbf{Detection and control of material phases with nonlinear approaches}

The emergence of many novel quantum phases is often revealed by discontinuities in thermodynamic probes, however the details of the order parameter can be obscured. Nonlinear techniques are proving very valuable in this regard; the use of optical SHG to uncover the electronic nematic phase in the metallic pyrcholore Cd$_{2}$Re$_{2}$O$_{7}$ is a recent example~\cite{Harter2017SHG}. The selection rules inherent to SHG were crucial to identifying the electronic origin of the new phase as well as the importance of nematic fluctuations in the material's structural transition.  SHG also confirmed the presence antiferromagnetism, imaged the domains and revealed the importance of the structure in exfoliated CrI$_{3}$ \cite{Sun2019SHG}. 

Beyond detection, nonlinear techniques have also begun to enable the emergence and control of new quantum states. For example, the use of circularly polarized light to control the gyrotropic order in the charge density wave state of TiSe$_{2}$, which was detected via CPGE \cite{Xu2020TiSe2}. Similarly, multi-pulse, intense THz beams with controled envelopes created a nonequilibrium gapless superconducting state in Nb$_{3}$Sn, which was detected via high harmonic generation \cite{Yang2019Nb3Sn}. 

In addition to correlated states, topological phases were controlled using intense THz pulses to induce a structural transition in WTe$_{2}$, observed via SHG \cite{Sie2019WTe2}. A more direct modification of the electronic structure via nonlinear electromagnetics was achieved in graphene using the Floquet mechanism to open a gap at the Dirac point \cite{mciver2020light}. Due to the induced Chern number and spin degeneracy, a plateau in the Hall conductance close to 2$e^{2}/h$ was observed when the Fermi level was placed into the gap. 

\section{V. Future Directions: correlations and topology}
So far, nonlinear approaches are mostly revealing symmetry and quantum geometrical properties rather than topology. The measurement of a quantized nonlinear response would be an undeniable sign of a topological state. Chiral topological semimetals (with either Weyl or multifold fermions) are promising in this regard, because they are predicted to display a quantized injection current~\cite{de2017quantized}. Because of the absence of chiral Weyl semimetal samples the focus has shifted to the chiral multifold metals RhSi~\cite{rees2019quantized} and CoSi~\cite{Ni2020b}, that show promising signs in ultrafast experiments aligned with material-specific predictions~\cite{chang2017unconventional,deJuan:2020jm,Ni2020b}. CoSi has a longer carrier lifetime which is an important requirement to observe quantization as scattering corrections are non-universal~\cite{de2017quantized,Konig17}. The experiment is also challenging because it is the current and not its generation rate (injection) which are typically measured. The effect of electron-electron interactions, which can affect non-linear responses through excitonic~\cite{Fei2020b,Chan2019} and other effects~\cite{Kishida:2000cv,Liu2017HHG,Silva:2018co}, also corrects quantization~\cite{avdoshkin2020interactions,Mandal_2020,Grushin2020}. 

Moving forward, nonlinear approaches provide a road map to reveal, create, and control novel quantum materials and states (Fig.~\ref{fig:overview}). Multiferroics break both inversion and time reversal symmetry, often supporting skyrmions, a vortex-like topological object in real space~\cite{seki2012observation}. Nonlinear probes can provide new insights into its Berry curvature, topology and magnetoelectric coupling. Moreover, one can employ high intensity beams to manipulate the lattice and thus magnetic interactions to convert a frustrated magnet into a true topological spin liquid. As suggested theoretically,\cite{TOka2019FloqRev} by employing a Floquet approach one can further tune the system to a non-Abelian phase and use local beams to move and ultimately braid the non-Abelian excitations. This achievement is crucial to understanding their non-trivial statistics and for their eventual use in a future topological quantum computer. Given recent advances in employing spontaneous Raman to measure spin liquids~\cite{Wang2020KitRam}, it would be highly desirable to adapt well established nonlinear Raman approaches (e.g. Coherent anti-Stokes Raman or Stimulated Raman Spectroscopy)~\cite{boyd2020nonlinear} to either measure or control the low energy excitations in these magnetic topological systems. Indeed this suggests ultimate control will require finding the right materials, devices, (Table.~\ref{Fig:3Dmat} and ~\ref{Fig:2Dmat}) , and new tools with local and non-local optical manipulation. One example is the recent demonstrations of gate tunable, nanoscale, broadband nonlinear detection and generation with graphene on re-writeable LAO/STO~\cite{sheridan2019gatetunable}.  It is interesting to note such advances may also answer other exciting scientific questions. For example, how to image and manipulate individual domains and/or the quantum geometry in momentum space? Similarly can we explore the higher order distribution of the Berry curvature? 

Improved material control will also be central to disentangling the role of disorder and other scattering mechanisms (e.g., electron-phonon coupling) in the various nonlinear responses. Such advances will rely on new theoretical directions to understand how to achieve ultimate control of quantum materials and utilize nonlinear electromagnetism to reveal new physical phenomena, for example, uncovering robust quantization and fractionalization of quasi-particles. Crucial to such efforts is the combination of the nonlinear techniques outlined here with conventional methods of material manipulation (electrostatic gating, pressure, magnetic fields). Such combinations are important to distinguishing the intrinsic effect of quantum geometry from other contributions (e.g., impurity, thermal effect) (Box 1 and Box 2), as well as the fine tuning needed to uncover emergent phases. This will require producing ever higher intensity, lower energy and shorter pulses. To this end, we envision utilizing metamaterial, plasmonic, polaronic, or photonic approaches with novel heterostructures of 2D and 3D materials. Near-field techniques are also likely to play an increasing role in such efforts, while simultaneously offering access to boundary modes and finite momentum responses. The rapid pace of advances in ultrafast optics, fabrication techniques, new materials and theoretical approaches will provide many opportunities to uncover, probe and control quantum materials. 

\vspace{0.5cm}

\vspace{0.5cm}
\noindent \textbf{Acknowledgement:} We thank F. de Juan, J. Song, S. Xu, and Y. Zhang for discussions and critical reading of the manuscript. K.S.B. is grateful for the primary support of the US Department of Energy (DOE), Office of Science, Office of Basic Energy Sciences under award no. DE-SC0018675. Q.M. is supported by the Center for the Advancement of Topological Semimetals, an Energy Frontier Research Center funded by the U.S. Department of Energy Office of Science, through the Ames Laboratory under contract DE-AC02-07CH11358 (manuscript preparation and writing). A.G.G. is supported by the ANR under the grant ANR-18-CE30-0001-01 and by the European Union Horizon 2020 research and innovation program under grant agreement No.~829044~(SCHINES).

\vspace{0.1cm}
\noindent \textbf{Author contributions:} K.S.B. conceived of the project. All authors wrote the manuscript together. 

\vspace{0.1cm}
\noindent \textbf{Data availability:} Correspondence and requests for materials should be addressed to K.S. Burch ~(email: ks.burch@bc.edu).

\vspace{0.1cm}
\noindent \textbf{Competing financial interests:} The authors declare no competing financial interests.

\bibliographystyle{naturemag}
\bibliography{Nonlinear_refs_Resubmit.bib}

\begin{widetext}

\newpage

\section{Box 1: Practical aspects in calculating nonlinear responses}\vspace{-2ex}
\noindent 
\textbf{Length gauge versus velocity gauge:}
Two gauges are primarily used in calculating nonlinear optical responses: the velocity gauge, based on the substitution $H(\mathbf{k})\to H(\mathbf{k}+\frac{e}{\hbar c}\mathbf{A})$~\cite{Ventura17,Parker2019diag}, and the position gauge, based on the coupling $\mathbf{r}\cdot \mathbf{E}$~\cite{Sipe00,Ventura17}. The gauges are equivalent as they are related by a unitary transformation  (e.g., Refs.~\cite{Parker2019diag,Ventura17}). However, in practice there are benefits and drawbacks to each (see Table~\ref{Table:gauge}), adapted from \cite{Parker2019diag,Ventura17,deJuan:2020jm}). An important difference between the two gauges lead to unphysical approximations. The shift current in the position gauge is written in terms of matrix elements that involve a maximum of two bands, while in the velocity gauge they involve sums over multiple bands. Although they are equivalent (via sum rules~\cite{Sipe00}) often one wants to calculate the response of an effective Hamiltonian involving a small subset of bands. It can be tempting to truncate the summation of the bands in the velocity gauge, keeping only terms between the bands of interest (e.g. the two bands of a Weyl node). However, this leads to incorrect results in the velocity gauge~\cite{Ventura17}. In general, whether a particular response coefficient involves a certain number of bands is a gauge dependent statement, and hence is not physical~\cite{Parker2019diag}. 

\begin{table}[H]
\vspace{-2ex}
{\small
\begin{tabular}{|p{0.06\textwidth}@{\hspace{0.02\textwidth}}|p{0.44\textwidth}@{\hspace{0.02\textwidth}}|p{0.44\textwidth}|}\hline \textbf{Gauge} & \textbf{Advantages} & \textbf{Disadvantages} \\\hline 
Velocity \newline ($\vec{v}\cdot\vec{A}$) & 
\tabitem Easy Feynman diagramatic formulation

\tabitem Suitable numerically: no $\vec{k}$ derivatives of position operator &
\tabitem Truncations of band summations can lead to unphysical result

\tabitem Important to handle unphysical low frequency divergences\\\hline
Length \newline ($\vec{r}\cdot\vec{E}$) &
\tabitem Easier link to semiclassical contributions

\tabitem Certain topological responses are transparent &
\tabitem $\vec{k}$ derivatives of position operator not suitable numerically

\tabitem Higher order frequency poles must be dealt with\\\hline
\end{tabular}\vspace{-1ex}
\caption{\textbf{Comparison of length and velocity gauges}}
\label{Table:gauge}}
\vspace{-3ex}
\end{table}

\noindent
\textbf{Tight-binding approximation:} The comparison between first principles and tight-binding calculations of nonlinear responses can be subtle. Indeed, the tight-binding approximation implicitly discards off-diagonal position matrix elements~\cite{Azpiroz2018}. Neglecting these may lead to sizable errors in the predicted shift current, even when the band dispersion and dielectric function are correct.\\

\noindent
\textbf{Spatially varying probes:} The beam's spatial dependence is described by responses at finite $\mathbf{q}$, the photon momentum. One example is the photon-drag effect where the transfer of the photon's linear momentum to the excited electron induces a current, sometimes explained by momentum dependent scattering~\cite{Glazov:2014cp,PhysRevB.93.125434}. The spatial intensity gradient of the beam can also generate a photocurrent~\cite{ji2019spatially}. This spatial photogalvanic effect, lowers the symmetry of the response resulting in currents oriented along directions that would be prohibited by crystalline symmetries. If the beam inhomogeneity produces an orbital angular momentum (OAM), a photocurrent proportional to the OAM can be generated~\cite{ji2020photocurrent}. The dependence on OAM results from Berry curvature and scattering that is asymmetric in strength with respect to exchange of the initial and final state~\cite{ji2019spatially,ji2020photocurrent}.\\

\noindent
\textbf{Time modulations:} Effects in topological phases due to short-time light pulses are starting to be addressed\cite{Moessner:2017ur,TOka2019FloqRev,luo2019ultrafast,sirica2020photocurrent}. For example, if the duration of the probing pulse is longer than the scattering time, the injection current can saturate to a non-universal value which depends on the scattering time. Since the injection current is sensitive to topological properties, this instance must be taken into account in the interpretation of experiments in search for a quantized CPGE~\cite{Konig17,rees2019quantized,Ni2020b}. A significant example a periodic pumps that can create dynamical states, which only exist out of equilibrium and may have different topology and symmetry~\cite{Moessner:2017ur}. This is described using Floquet theory, the time-domain analogue of Bloch's theorem. Specifically, we employ Floquet states that repeat in time with quasi-energies defined modulo the frequency. Floquet states can be viewed as nonlinear in the sense that the response to an external field is treated nonperturbatively (strong light-matter coupling regime). Upon expanding in powers of the external fields, it is possible to recover the different $\sigma^{(i)}$~\cite{Morimoto:2016iu}.\\

\noindent
\textbf{Band Structure:} Nonlinearities can arise from non-quadratic band dispersion, or high-field induced density distribution.\cite{SturmanBook,Glazov:2014cp,Gullans2013Nonlin,Yoshikawa2017HHG,boyd2020nonlinear, SturmanBook} For Dirac dispersions, velocity only depends on the direction of the momentum. Since the electron momentum follows the AC field, the velocity (i.e., current) is: $v(t)\propto \pm \mathrm{sign}(\sin(\omega t))=\pm\frac{4}{\pi}(\sin(\omega t)+(1/3)\sin(3\omega t)+(1/5)\sin(5\omega t)+...)$. As the higher-order terms are relatively close to the linear one, linear dispersions enable sizeable, odd-power nonlinear responses~\cite{Glazov:2014cp,Gullans2013Nonlin}. Band dispersion is also crucial when nonlinearity is governed by quantum geometry~\cite{Sipe00,zhang2018photogalvanic}. For example, the Berry monopole of a Weyl semimetal implies nonlinear responses diverge as $\omega \rightarrow 0$, but the vanishing density of states places the maximum at finite frequency.\cite{ahn2020low,Yang17} The resonance conditions in Harmonic generation, Floquet and photocurrent experiments are also sensitive to Fermi level~\cite{mciver2020light,Mikhailov2014THG, Glazov:2014cp,Pan2017CPGE}.\\

\noindent
\textbf{Scattering effects:} The role of scattering in nonlinear responses has a long history\cite{BelinicherTransient}, but remains an active open question~\cite{Sotome2019Shift,Konig17,Golub18,Nandy2019,Konig19,isobe2020high,Du:2019vg}. For example the ballistic contribution~\cite{SturmanBook} or low energy photocurrents\cite{Glazov:2014cp} that can arise from asymmetric scattering of photo-excited carriers. In the nonlinear Hall effect, contributions from skew-scattering and side-jump processes have been calculated~\cite{Nandy2019,Konig19,Du:2019vg,xiao2019theory,isobe2020high}. Skew-scattering can dominate the nonlinear response, especially if the Berry curvature dipole is zero (e.g. in hexagonal materials, see: Fig.~\ref{fig:Berry2D}\textbf{b}\&\textbf{e})~\cite{Nandy2019,Konig19,Du:2019vg,isobe2020high}. Another example is the free-carrier contribution to difference-frequency response,\cite{deJuan:2020jm} which is well defined in the clean limit ($\omega\tau\to\infty$), yet its survival when disorder dominates is unclear~\cite{BelinicherTransient}. For shift currents, scattering and the stimulated recombination of photoexcited carriers must be included~\cite{Sotome2019Shift}.
\newpage

\section{Box 2: Experimental Complications}
In planning and interpreting nonlinear experiments, a variety of contributions other than the the quantum geometry or crystal symmetry must be carefully considered. A summary is listed in the Table \ref{Table:Other}.

\begin{table}[H]
    \centering
\begin{tabular}{|p{0.34\textwidth} m{0.01\textwidth}|p{0.3\textwidth}m{0.01\textwidth}|p{0.3\textwidth}|}\hline\textbf{Cause} && \textbf{Experimental Considerations} && \textbf{Experimental Signatures} \\\hline \uline{Photo-Thermal}: Anisotropic heating from the laser causes a thermal gradient, which  produces a photovoltage via the Seebeck Effect. && Beam Size \& Homogeneity\newline Device Dimensions\newline Contact Uniformity \newline Relative Alignment of Laser \& Active Area && Photovoltage reverses as laser is scanned.\newline Photovoltage reverses from hole to electron dominated transport.\newline Wavelength/polarization dependence proportional only to absorption.\\\hline
\uline{Photo-Gating:} Increase in carrier density due to defects trapping of one of the photoexcited charges. && Intensity dependence \newline Material Uniformity \newline  && Photoconductivity that saturates with intensity or changes in gate voltage dependence of resistance.\\\hline
\uline{Built-in Electric Fields:} large work function difference in with the contact produces a Schottky barrier or local doping, that can break symmetry or produce photocurrents. && Contact material \newline Contact attachment technique \newline Spatial dependence of signal \newline Spatial layout of contacts  && Enhanced signal near contacts\newline Response not allowed by symmetry \newline Nonlinear dc current versus voltage. \\\hline
\uline{Optical Artifacts:} Wavelength, temperature, spatial or polarization dependence resulting from absorption/reflection of sample and imperfect optical elements. && Separately measure optical constants \newline Correct for losses \newline Lenses instead of mirrors \newline Photoelastic modulators && Polarization Dependent Absorption \newline Beam ``Dancing'' (waveplate parallelism) \newline Different spatial dependence of reflectance and photocurrent\\\hline
\end{tabular}
\caption{\textbf{Summary of Additional Considerations in Nonlinear Experiments}}
\label{Table:Other}
\end{table}

\noindent
\textbf{Optical Considerations:} Inhomogeniety in the laser, or its motion on the sample can induce photothermal responses, as the absorbed power leads to temperature gradients. Such effects can arise in a variety of ways, including the laser not being normal to the device surface or non-uniform optical elements. The motion of the beam inherent to quarter and half wave plates is particularly challenging. Since the front and back surfaces are never perfectly perpendicular, as the element is rotated the laser's pointing vector moves resulting in some offset of the spot. Combined with photothermal or contact effects this can appear as a linear or circular photogalvanic effect~\cite{osterhoudt2019colossal}. The optical elements, angle of incidence and photothermal effects can also conspire to produce artifacts. For example, the quarter and half wave plates rely on anisotropic index of refraction, that will also produce small modulations in the intensity upon rotation. Similarly, the same broken symmetries that allow nonlinear terms also typically produce anisotropy in the optical constants. As such the absorption/reflection and photothermal response will be modulated as the polarization state is changed. However this is primarily a concern for photocurrent measurements of the linear polarization and typically does not appear in circular terms. Similarly, the change in absorption/reflection with wavelength must be accounted for when comparing nonlinear responses measured at different laser wavelengths. These concerns can be mitigated through a number of approaches, including large laser spots relative to the device and accounting for changes due to the optical response of the material in the analysis~\cite{osterhoudt2019colossal,Glazov:2014cp}. 

\noindent
\textbf{Contacts:} In photocurrent measurements, contacts play a crucial and sometimes nontrivial role. Depending on the relative band alignment and work functions, the metal contact can locally dope or produce a Shotkky barrier. Both effects tend to be spatially inhomogeneous, and as such can break the local symmetry~\cite{LiuNonlinDetect}. While this can be employed to enable nonlinear responses not originally possible due to symmetry constraints~\cite{Dhara:2015di}, the built-in potentials can also result in strong photocurrents near the contacts. Similarly, the change in Seebeck coefficient between the sample and metal contact can similarly produce a a strong photothermal response. Depending on the distance between the contacts and the spot size, such effects can appear to be from the device itself. The plasmonic and edge responses of the contacts can produce local and polarization dependent enhancements that can mimic the polarization dependence expected from the intrinsic nonlinear photocurrents. Thus care must be taken in choosing the contact material, shape, and size relative to the beam. Furthermore the spatial and polarization dependence of the response can help to identify and eliminate these extrinsic contributions.

\noindent
\textbf{Inhomogeneity:} Structural defects or domains (e.g. magnetic, ferroelectric) can locally break the symmetry producing even harmonic generation or photocurrents~\cite{Harter2017SHG}. This can be especially important in materials where the domain size or disorder is on length scales far below the light wavelength. Impurities and structural defects can also change the local chemical potential and/or add levels that act as traps for carriers. The former will alter the local Seebeck coefficient, producing photothermal signals. The traps are unlikely to produce photocurrent or second harmonic signals by themselves, but may introduce additional optical resonances that modify the wavelength dependence. In addition, the defects can trap photoexcited carriers, producing an increase in the carrier population of the opposite charge. This photo-gating effect is sometimes used for photodetection and is also unlikely to produce a nonlinear signal, but will modify the conductivity and thus can result in changes to the photocurrrent produced as well as the intensity dependence~\cite{LiuNonlinDetect}.
\newpage

\section{Box 3: Topological bulk materials}

Among the many topological materials discovered in 3D, Weyl semimetals~\cite{armitage2018weyl} have been the mostly intensively studied in the context of quantum geometry and nonlinear responses. Indeed, a Weyl semimetal phase can only be realized in non-centrosymmetric or magnetic materials, which is exactly the symmetry constraint that allows a non-zero Berry curvature. Moreover, the Berry curvature is diverging at the Weyl nodes, suggesting that significant responses can arise from the quantum geometrical properties. As a representative for inversion-breaking Weyl semimetals, the class of TaAs has shown optical SHG, CPGE and shift current as discussed in Section IV. In addition, TaAs is polar, enabling the existence of Berry curvature dipole that can lead to a second-order anomalous Hall effect~\cite{moore2010confinement,sodemann2015quantum}. In general, Weyl nodes occur at generic $k$ points. Without additional symmetries, the Weyl cones are always tilted along a certain momentum direction. The degree of tilt leads to the classification of type I and type II Weyl nodes. In type I Weyl semimetals, such as the TaAs family, the Weyl cone is mildly tilted with a standard point-like Fermi surface at the Weyl crossing. In contrast, type II Weyl semimetals, such as LaAlGe, $T_\mathrm{d}$ - Mo(W)Te$_2$, TaIrTe$_4$~\cite{armitage2018weyl,soluyanov2015type}, feature significantly tilted Weyl cones with finite electron and hole pockets at the Weyl crossing. The Weyl cone tilting in both type I and II Weyl semimetals is critical for CPGE generation as further elaborated in Section IV.  

\setcounter{figure}{2}
\renewcommand{\figurename}{TABLE}

\begin{wrapfigure}{r}{0.6\textwidth}
    \includegraphics[width=0.6\textwidth]{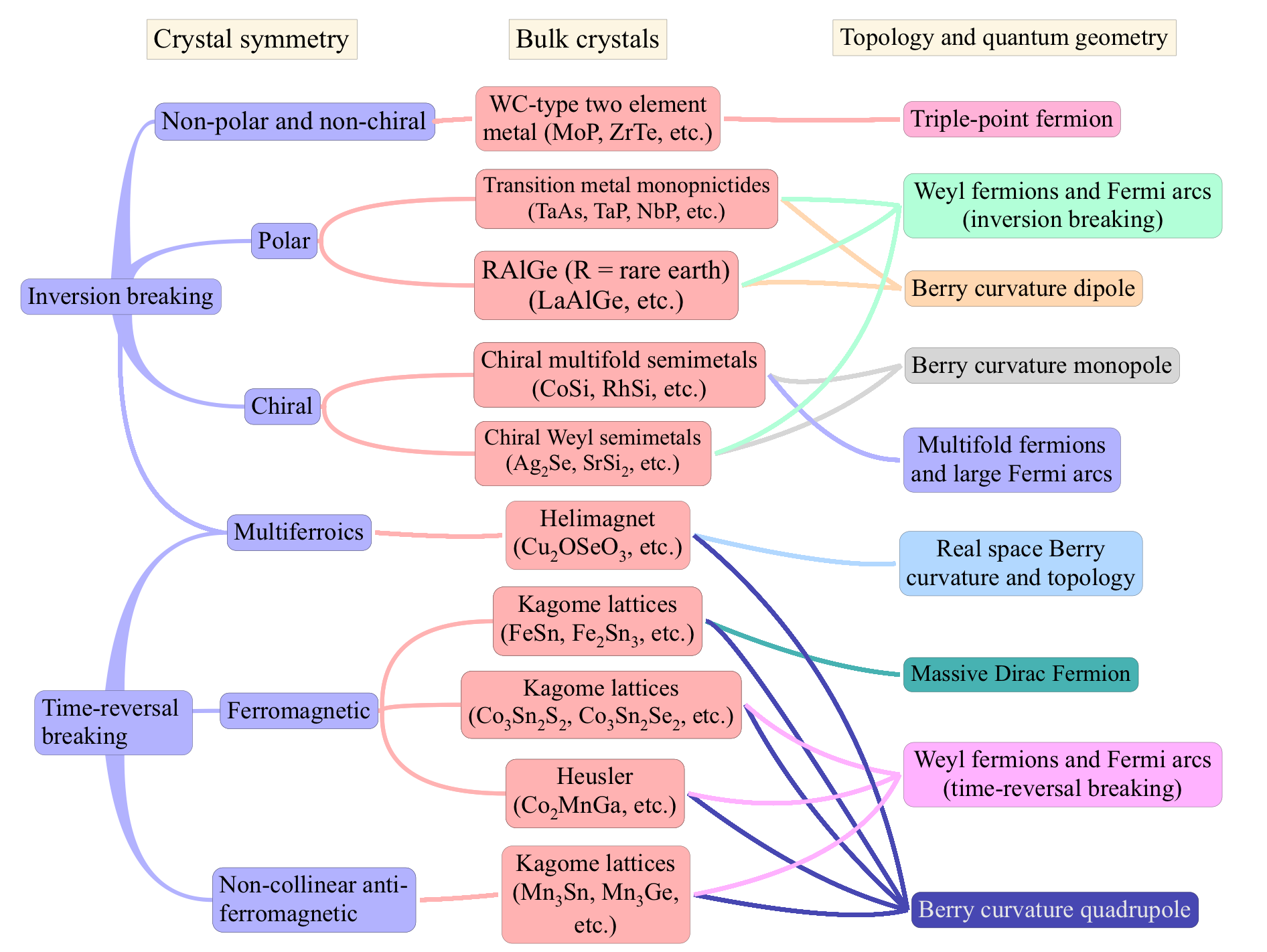}
\caption{\textbf{Quantum materials with important topological and quantum geometrical properties in its bulk form.}
}
\label{Fig:3Dmat}
\end{wrapfigure}

Recently, three groups independently identified ferromagnetic crystals Co$_2$MnGa and Co$_3$Sn$_2$S$_2$ as magnetic Weyl semimetals with spectroscopic (ARPES and STM) experiments~\cite{belopolski2019discovery, liu2019magnetic, morali2019fermi}. Moreover, anti-ferromagnetic materials Mn$_3$Ge and Mn$_3$Sn with noncollinear spin texture are predicted to be magnetic Weyl semimetals (Table~\ref{Fig:3Dmat}), with experimentally observed large linear anomalous Hall responses~\cite{kuroda2017evidence}. All these magnetic Weyl materials are ideal candidates to realize the recent prediction that the Berry curvature quadruple could lead to strong third-harmonic nonlinear responses~\cite{Parker2019diag}. 

Chiral crystals, a subgroup of non-centrosymmetric systems, are materials with a well-defined handedness due to the lack of mirror symmetries. Theory predicts that non-magnetic chiral crystals always host either Weyl or multifold nodes~\cite{Bradlyn2016,chang2017unconventional,tang2017multiple,chang2018topological} (Table~\ref{Fig:3Dmat}). They feature (1) nodes located at Kramers or time-reversal-invariant-momenta (TRIM) points, (2) a node chirality that is directly determined by the chirality of the lattice, (3) nodes of opposite chirality at different energies due to the low crystal symmetry, and (4) Fermi arc surface states that can span over the entire Brillouin zone, with multiple material candidates\cite{huang2016new,Schroter:2019du,chang2017unconventional,tang2017multiple,sanchez2019topological,Rao:2019ha,Takane19}. The chiral lattice allows distinct nonlinear responses beyond those in non-chiral Weyl semimetals, such as the quantized CPGE \cite{de2017quantized}. In these materials the absence of mirror symmetry significantly separates opposite chiralities in energy and allows only one node to be optically activated, quantizing the CPGE~\cite{chang2017unconventional,Flicker2018Chiral} (see also Section II).

In addition, the extended Fermi arc surface states in chiral semimetals have been suggested to contribute to non-linear photocurrents. Usually the bulk and surface responses are entangled and often the bulk response dominates. Theory predicts that the injection current generated from the surface Fermi arc states can be perpendicular to that generated in the bulk regardless of the choice of material surfaces in the chiral crystals RhSi and CoSi \cite{chang2020unconventional,Flicker2018Chiral}. 
\newpage

\section{Box 4: Van der Waals layered materials}

Van der Waals (vdW) thin layers and their heterostructures represent a highly tunable material platform allowing for advanced device fabrication, manipulation, and control methods \cite{yankowitz2019van, song2018electron}. Specifically, their chemical potential can be controlled by electrostatic gating; their symmetry, electronic structures, and quantum phases can be dramatically modified by layer stacking and twisting, and by strain engineering; they can be coupled with superconductors or ferromagnets for proximity effects and with metamaterials, cavity, photonic or plasmonic structures for enhanced light-matter interactions. Compared with bulk crystals, their versatility offers freedom to explore topological and Berry curvature physics in a highly controllable way.

\begin{wrapfigure}{r}{0.55\textwidth}
    \includegraphics[width=0.55\textwidth]{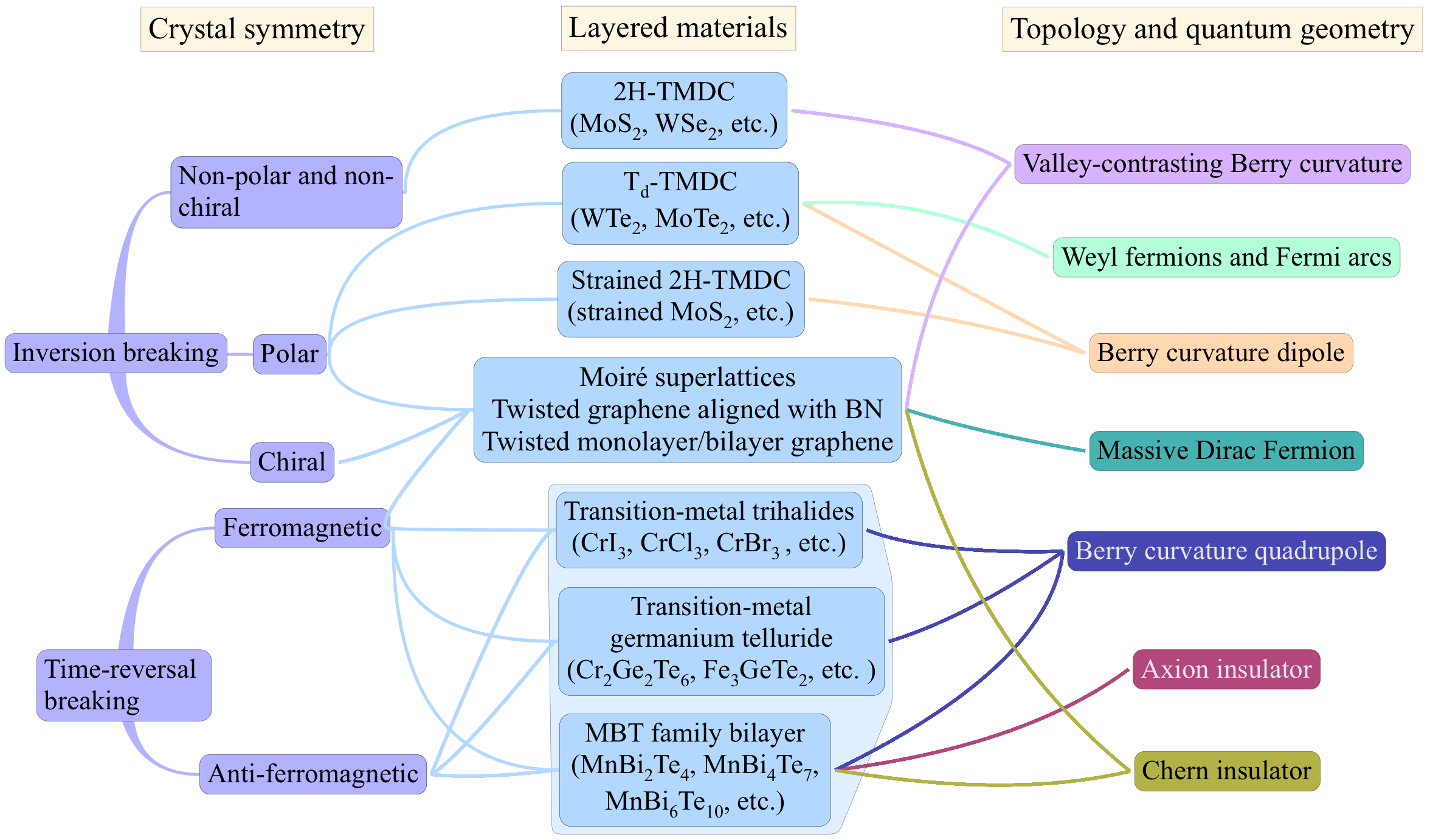}
\caption{\textbf{Layered materials with important topological and quantum geometrical properties.}
}
\label{Fig:2Dmat}
\end{wrapfigure}

The canonical example is pristine graphene with both inversion and time-reversal symmetry. These symmetry properties lead to identically vanishing Berry curvature in momentum space~\cite{sarma2011electronic}. However, one can break the inversion symmetry by crystallographically aligning the monolayer with hexagonal boron nitride (hBN) or by applying a perpendicular electric field to a bilayer~\cite{yankowitz2019van}. A band gap then opens up at charge neutrality with strong Berry curvature concentrated near band edges and opposite between K and K' valleys (Fig.~\ref{fig:Berry2D}\textbf{a}). Such valley-contrasting Berry curvature also exists in semiconducting ($H$ phase) transition metal dichalcogenides or TMDCs (e.g., MoS$_2$). $T$ phase TMDCs (e.g., W(Mo)Te$_2$) hosts different topological and geometrical properties from the $H$ phase. Particularly, monolayer 1$T$'- WTe$_2$ has is a 2D topolgical insulator with inversion symmetry~\cite{fei2017edge, wu2018observation}. One can break the inversion symmetry by applying a perpendicular electric field and probe nonlinear responses arising from the topological bands~\cite{xu2018electrically}. Bilayer WTe$_2$ is strongly inversion-broken due to the stacking order and features both out-of-plane and in-plane polarization. Particularly, the in-plane polarization leads to a dipolar distribution of Berry curvature for realizing nonlinear anomalous Hall~\cite{moore2010confinement, sodemann2015quantum}. 
 
In layered magnets, the symmetry, quantum geometry and topology are intimately correlated with layer-resolved spin structures~\cite{BurchNatMag18}. The first examples include semiconducting transition metal trihalides~\cite{huang2017layer} and transition metal germanium telluride~\cite{gong2017discovery,deng2018gate} (Table~\ref{Fig:2Dmat}). One can control the interlayer spin configuration (aligned or anti-aligned) by applying magnetic field, electric field, and hydrostatic pressure~\cite{BurchNatMag18}. The ability to control interlayer spin textures can directly translate into the control of spin-related nonlinear responses~\cite{Sun2019SHG,Holder19}. More recently, (Bi$_2$Te$_3$)$_m$(MnTe$_2$)$_n$~\cite{otrokov2019prediction} emerges as a 2D topological magnet (Table~\ref{Fig:2Dmat}). This class can be viewed as Bi$_2$Te$_3$ (a topological insulator) intercalated by MnTe$_2$ (a magnetic metal) at different stoichiometry ratios. Exfoliated thin layer of MnBi$_2$Te$_4$ shows quantized anomalous Hall conductance at zero magnetic field (for odd number of layers) and at finite magnetic field~\cite{deng2020quantum}. More interestingly, the even number of layers breaks the inversion symmetry due to the AFM ordering, and at the same time, is an Axion insulator~\cite{liu2020robust}.  These properties can result in exotic nonlinear responses~\cite{Fei2020} that await exploration both theoretically and experimentally. Similarly unexplored are the nonlinear responses due to the nontrivial topology of the spin sector of magnet 2D materials. For example the potential spin-liquid state in RuCl$_{3}$\cite{Wang2020KitRam}. Interestingly this material also offers extreme and local doping\cite{YWang2020ModDop} that could be exploited for symmetry breaking in heterostructures to induce nonlinear effects and novel topological phases. 

Another fruitful avenue is the vdW approach to create heterostructures with new symmetry and topological properties. Beyond the case of graphene on hBN mentioned above, an interesting example is the spin-orbit-driven band inversion realized in Bernal bilayer graphene after it is proximitized by an adjacent WSe$_2$ layer~\cite{island2019spin}. Such a band inversion could lead to a nontrivial Berry curvature texture. Most recently, the discovery of correlated insulating states and superconductivity in magic-angle moir\'e bilayer graphene~\cite{cao2018unconventional} triggered significant interest on twist moir\'e superlattices. 
These long-wavelength moir\'e systems not only produce flat electronic bands and van Hove singularities that are responsible for strong correlations~\cite{tang2020simulation,regan2020mott} but also enable effective symmetry engineering of the atomic lattices (Table~\ref{Fig:2Dmat}). For example, marginally-twisted MoSe$_2$-MoSe$_2$ form up-down polarization domains~\cite{sung2020broken}. Moreover, strong interaction effects can lead to spontaneous symmetry breaking, giving rise to new topological and quantum geometrical properties~\cite{balents2020superconductivity,zheng2020unconventional}. Lastly, these systems are offering unique opportunities for plasmonic and photonic engineering, with the potential to further enhance nonlinear responses~\cite{herzig2020rise}.

\newpage
\setcounter{figure}{0}
\renewcommand{\figurename}{Fig.}

\begin{figure}
    \centering
    \includegraphics[width=0.8\textwidth]{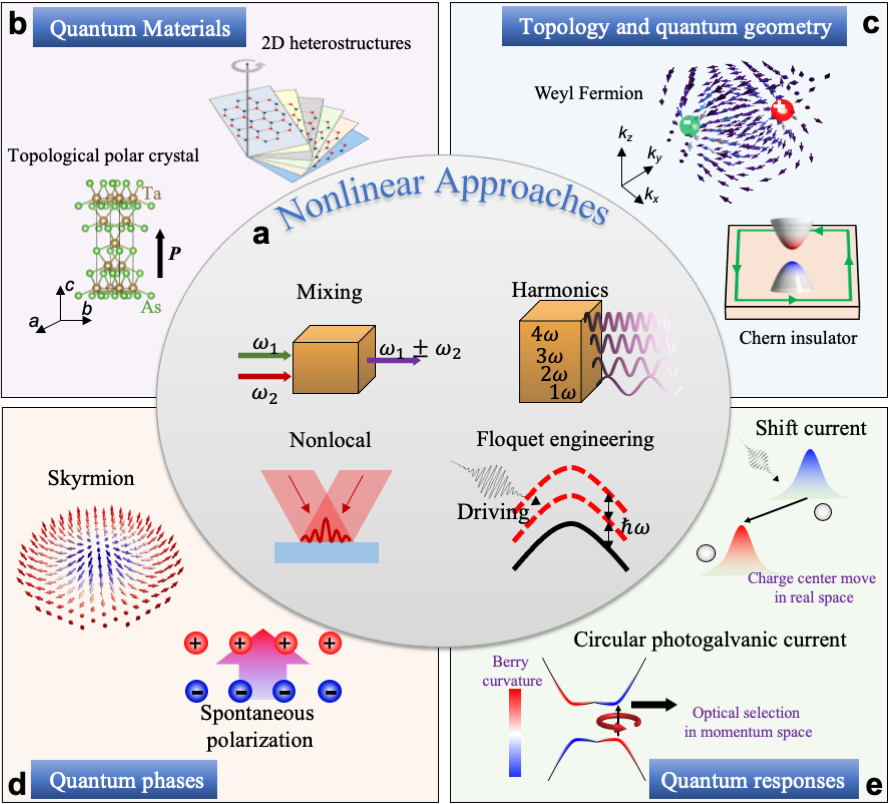}
    \caption{\textbf{Overview of nonlinear approaches and their applications in studying new materials and quantum phases, revealing topology and quantum geometry and detecting new electromagnetic responses.} \textbf{a,} This review covers experiments in both 3D bulk crystals and 2D atomically thin materials, such as topological Weyl semimetal TaAs and atomically thin 2D heterostructures. \textbf{b,} We focus on how to apply these nonlinear approaches to detect and manipulate topological and geometrical properties in quantum materials. \textbf{c,} Nonlinear responses can be used to probe emergent phases and broken symmetry states. \textbf{d,} In this review we highlight the shift current and the circular photogalvanic current as two representative second-order optical responses. In a simplified picture, shift current is generated due to an instantaneous shift of the center of electron wavefunctions in real space upon photo-excitation, which is described by Berry connection. The circular photogalvanic current is generated from asymmetric optical excitation in momentum space by circular polarized light. In the two-band limit, the optical selection rule is determined by the Berry curvature distribution in momentum space.}
    \label{fig:overview}
\end{figure}

\begin{figure}
    \centering
    \includegraphics[width=\textwidth]{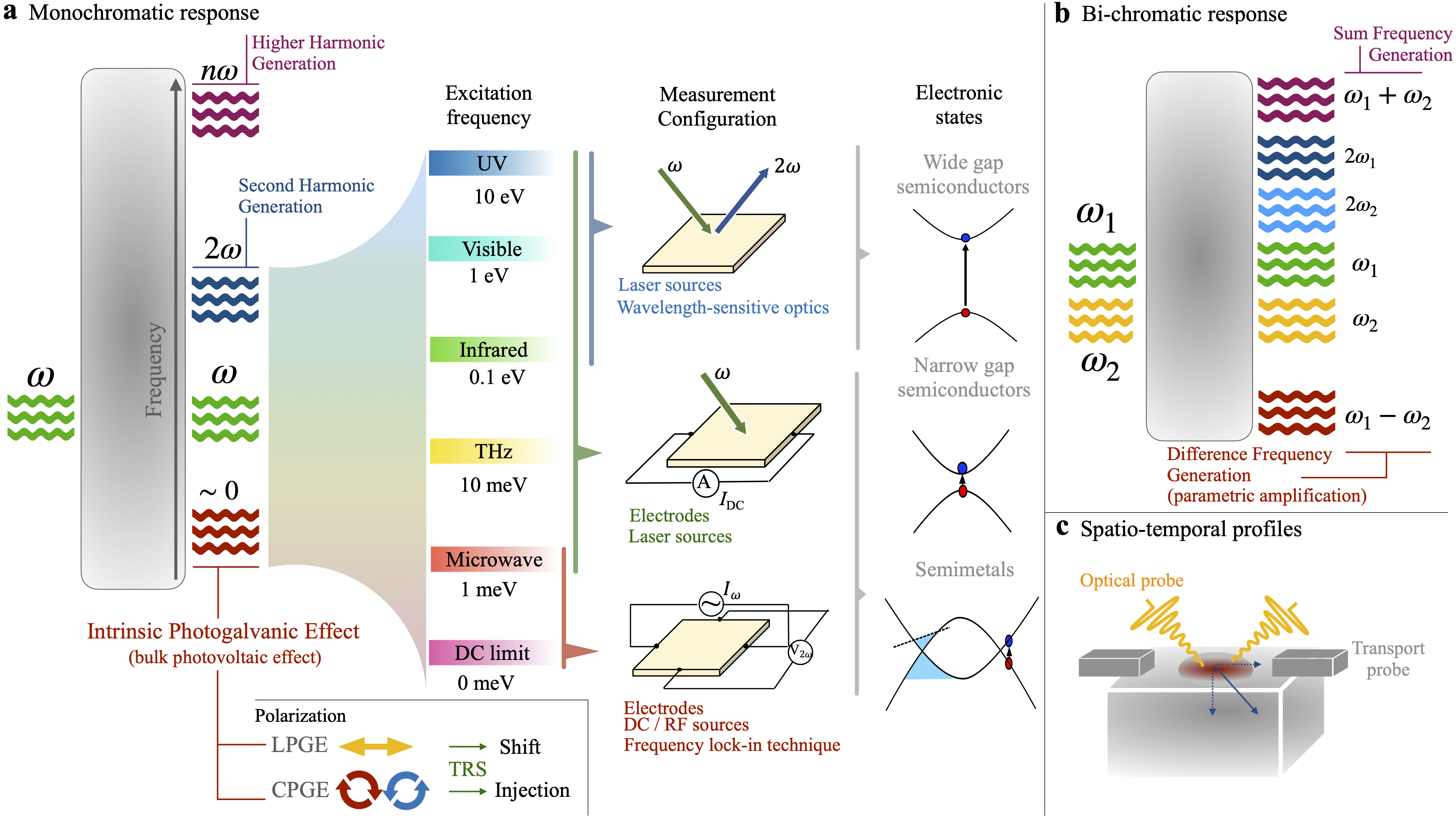}
    \caption{\textbf{Summary of second-order nonlinear processes}. \textbf{a,} When a continuous monochromatic beam of frequency $\omega$ is applied to crystal that lacks crystalline inversion symmetry it can generate second-harmonics at $2\omega$, and a DC photocurrent known as the intrinsic photogalvanic effect. The latter is known as the linear or circular photogalvanic effect for linear and circular incident polarizations, respectively. Depending on the frequency different probes, ranging from all electric to all-optical measurements can be performed. \textbf{b,} If two beams of frequencies $\omega_{1,2}$ are applied to the same crystal it can generate a response at the sum or difference of their frequencies, known as sum and difference frequency generation respectively, as well as responses that oscillates at $2\omega_{1,2}$ (second-harmonic generation). \textbf{c,} Nonlinear photocurrents can be detected experimentally by capturing the carriers with metallic contacts, or via an emitted pulse whose electric field that can be detected in the far field.}
    \label{fig:Possible}
\end{figure}

\begin{figure}
\centering
\includegraphics[width=\textwidth]{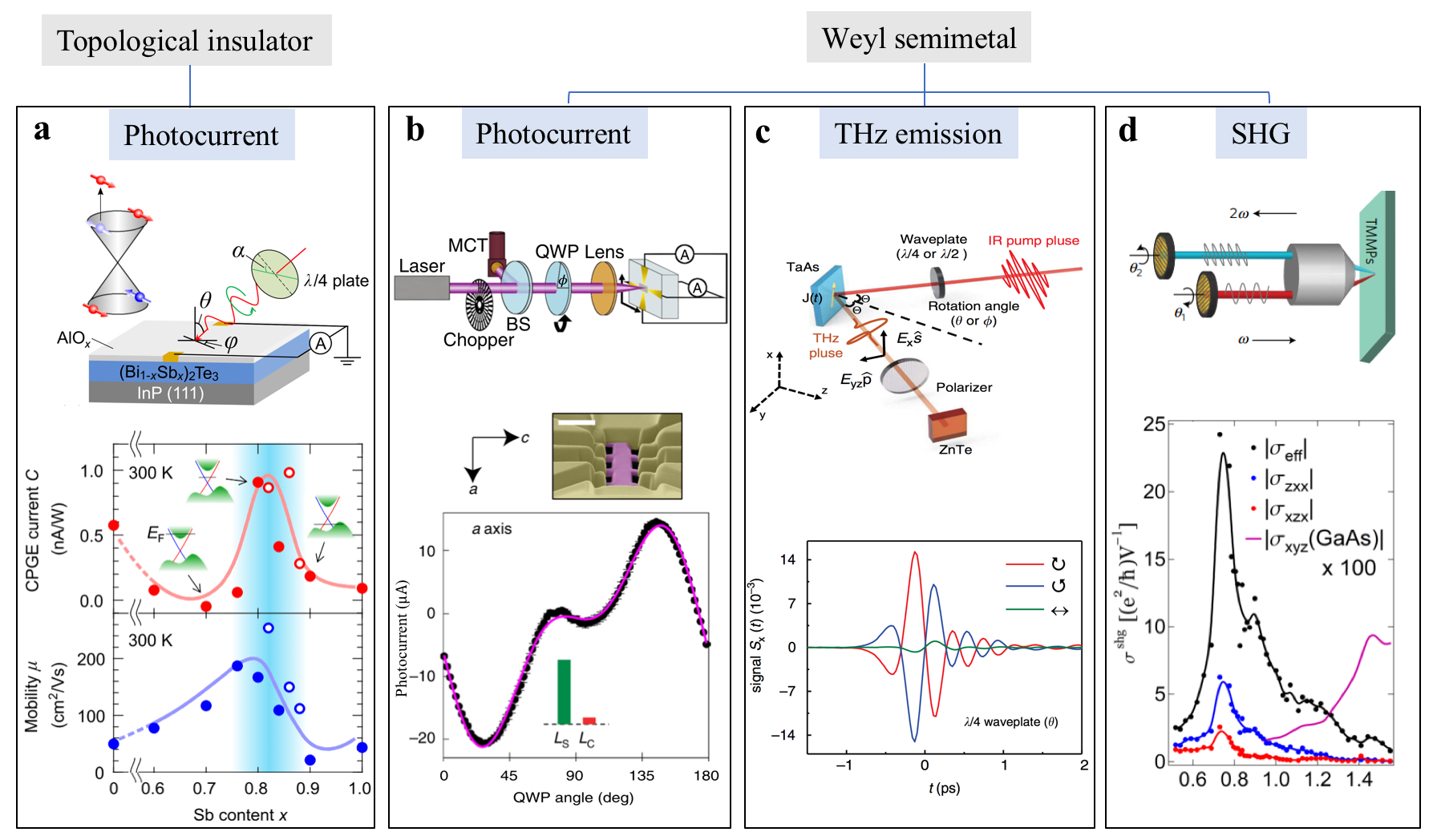}
\caption{\textbf{Nonlinear responses in topological insulators and inversion-breaking Weyl semimetals.} \textbf{a,} CPGE generation in Sb-doped Bi$_2$Te$_3$: experimental setup and the CPGE as a function of the doping level. The incident light has to be at an oblique angle to the surface to generate an in-plane CPGE due to the rotational symmetry about $z$ axis, adapted from \cite{2016PhRvB93h1403O}. \textbf{b,} Photocurrent generation in the topological polar crystal TaAs: experimental configuration, and the main observation for shift current. The DC photocurrent was generated by a continuous wave (CW) laser and measured through electrical contacts in a micro-sized device fabricated by focused ion beam (FIB). Scale bar: 5 $\mu$m. Adapted from Ref.~\cite{osterhoudt2019colossal}. \textbf{c,} THz emission experiment in TaAs: experimental configuration, and the main observation for CPGE. The photocurrent was generated by a femtosecond laser pulse and measured through emitted THz radiation in a non-contact manner and contains time-resolved information. This measurement setup is the key towards the observation of quantized CPGE~\cite{de2017quantized}. Adapted from Ref.~\cite{gao2020chiral}. \textbf {d,} SHG spectroscopy in TaAs, revealing a resonant peak around 0.7 eV. Adapted from Ref.~\cite{Wu2017giant,patankar2018resonance}.} 
\label{fig:3Dmat}
\end{figure}

\begin{figure}
    \centering
    \includegraphics[width=\textwidth]{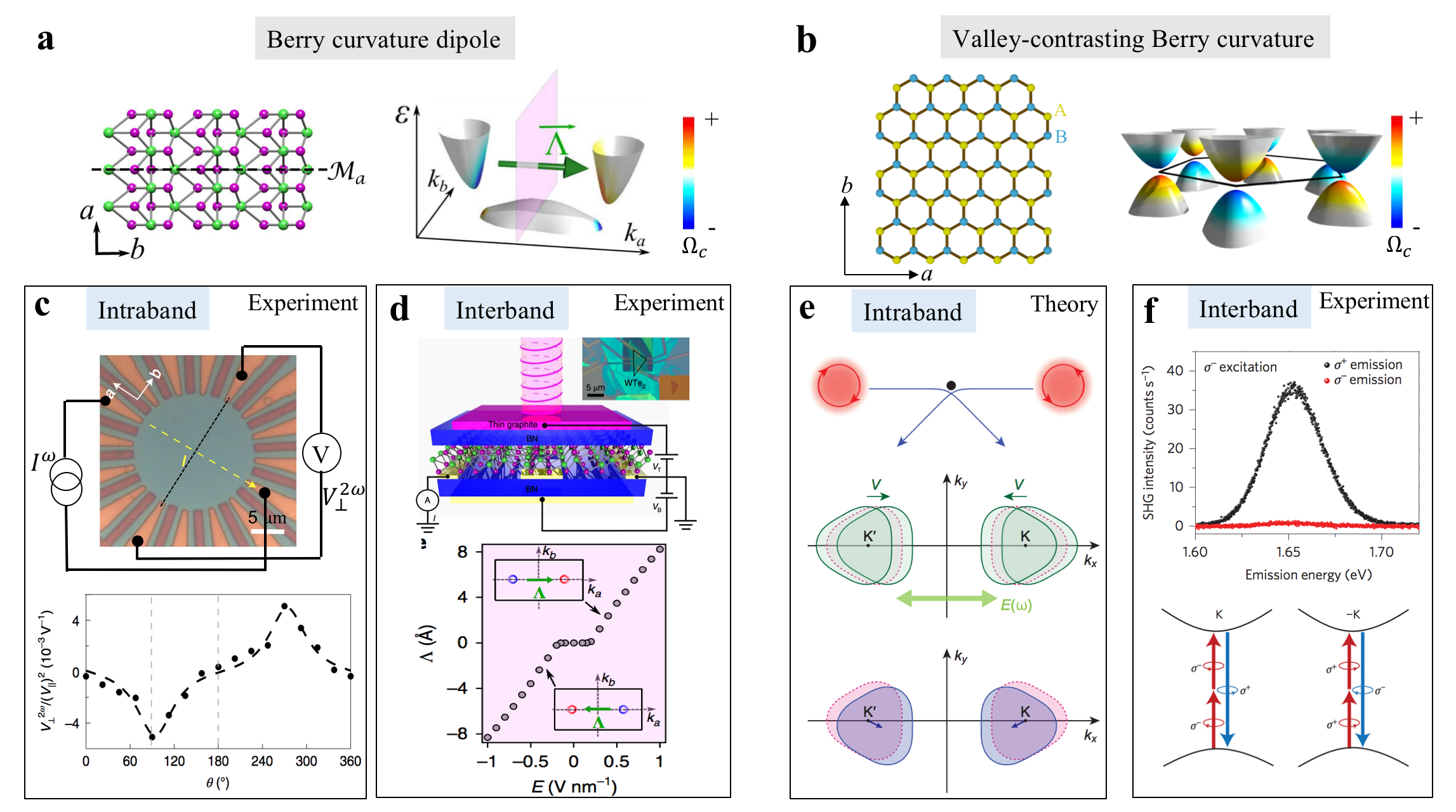}
    \caption{\textbf{Nonlinear responses enabled by Berry curvature dipole in non-centrosymmetric WTe$_2$ and valley-contrasting Berry curvature in non-centrosymmetric hexagonal lattices.} \textbf{a,} The lattice of WTe$_2$ and its band structure color-coded with Berry curvature, showing the dipole moment. \textbf{b,} The illustration of a hexagonal lattice with broken inversion and its band structure. The color bar represents the sign and strength of Berry curvature. \textbf{c,} The observation of nonlinear anomalous Hall effect as an intraband manifestation of the Berry curvature dipole effect. Adapted from Ref.~\cite{kang2019nonlinear}, also see Ref.~\cite{ma2019nonlinear}. \textbf{d,} The observation of in-plane CPGE with normal light incidence as an interband manifestation of the Berry curvature dipole effect.  Adapted from Ref.~\cite{xu2018electrically}. \textbf{e,} Second-order response due to skew scattering of chiral electrons. Adapted from Ref.~\cite{isobe2020high}. \textbf{f,} Optical second-harmonic generation in inversion broken WSe$_2$ with circular selection rules. Adapted from Ref.~\cite{seyler2015electrical}.}
    \label{fig:Berry2D}
\end{figure}

\end{widetext}

\end{document}